\pdfoutput=1

\documentclass[11pt,twoside,a4paper,cmspaper,final,collab]{cms-tdr}
\def\svnVersion{275802}\def\cmsCernNo{2013-037}\def\cmsCernDate{\today}\def\cmsMessage{Published in Physics Letters B as \href{http://dx.doi.org/10.1016/j.physletb/2015.01.034}{\doi{10.1016/j.physletb/2015.01.034.}}}

\begin{document}\cmsNoteHeader{HIN-14-002}

\hyphenation{had-ron-i-za-tion}
\hyphenation{cal-or-i-me-ter}
\hyphenation{de-vices}

\RCS$Revision: 270688 $
\RCS$HeadURL: svn+ssh://svn.cern.ch/reps/tdr2/papers/HIN-14-002/trunk/HIN-14-002.tex $
\RCS$Id: HIN-14-002.tex 270688 2014-12-08 12:32:42Z alverson $
\newlength\cmsFigWidth
\ifthenelse{\boolean{cms@external}}{\setlength\cmsFigWidth{0.85\columnwidth}}{\setlength\cmsFigWidth{0.4\textwidth}}
\ifthenelse{\boolean{cms@external}}{\providecommand{\cmsLeft}{top}}{\providecommand{\cmsLeft}{left}}
\ifthenelse{\boolean{cms@external}}{\providecommand{\cmsRight}{bottom}}{\providecommand{\cmsRight}{right}}
\newcommand {\roots}    {\ensuremath{\sqrt{s}}}
\newcommand {\rootsNN}  {\ensuremath{\sqrt{s_{_{NN}}}}}
\newcommand {\PbPb}  {\mbox{PbPb}}
\newcommand {\pPb}  {\ensuremath{\text{pPb}}\xspace}
\newcommand {\AuAu}  {\mbox{AuAu}}
\newcommand {\AonA}  {\ensuremath{\text{AA}}\xspace}
\newcommand {\vnsig}    {\ensuremath{v_{n}^\text{sig}}}
\newcommand {\vnbkg}    {\ensuremath{v_{n}^\text{bkg}}}
\newcommand {\vnobs}    {\ensuremath{v_{n}^\text{obs}}}
\newcommand {\vsecsig}    {\ensuremath{v_2^\text{sig}}}
\newcommand {\vsecbkg}    {\ensuremath{v_2^\text{bkg}}}
\newcommand {\vsecobs}    {\ensuremath{v_2^\text{obs}}}
\newcommand {\vtrdsig}    {\ensuremath{v_3^\text{sig}}}
\newcommand{\ket}       {\ensuremath{KE_{\mathrm{T}}}\xspace}
\newcommand {\deta}     {\ensuremath{\Delta\eta}}
\newcommand {\dphi}     {\ensuremath{\Delta\phi}}
\newcommand{\noff}    {\ensuremath{N_\text{trk}^\text{offline}}\xspace}
\newcommand {\dAu}    {\ensuremath{\text{dAu}}\xspace}
\newcommand {\ptass}       {\ensuremath{p_\mathrm{T}^{\text{assoc}}}}
\newcommand {\pttrg}       {\ensuremath{p_\mathrm{T}^{\text{trig}}}}
\newcommand {\ptref}       {\ensuremath{p_\mathrm{T}^{\text{ref}}}}
\providecommand{\EPOS} {\textsc{epos}\xspace}

\cmsNoteHeader{HIN-14-002} % This is over-written in the CMS environment: useful as preprint no. for export versions
\title{Long-range two-particle correlations of strange hadrons with charged particles in pPb and PbPb collisions at LHC energies}

\date{\today}

\abstract{
Measurements of two-particle angular correlations between an identified strange hadron ($\PKzS$ or $\PgL$/$\PagL$)
and a charged particle, emitted in pPb collisions, are presented over a wide
range in pseudorapidity and full azimuth. The data, corresponding to an integrated
luminosity of approximately 35\nbinv, were collected
at a nucleon-nucleon center-of-mass energy (\ensuremath{\sqrt{s_{_{NN}}}}) of 5.02\TeV with the CMS detector at the LHC. The results
are compared to semi-peripheral PbPb collision data at \ensuremath{\sqrt{s_{_{NN}}}} = 2.76\TeV,
covering similar charged-particle multiplicities in the events. The observed azimuthal
correlations at large relative pseudorapidity are used to extract the second-order ($v_2$) and third-order ($v_3$) anisotropy
harmonics of $\PKzS$ and $\PgL$/$\PagL$ particles.
These quantities are studied as a function of the charged-particle multiplicity in the event and
the transverse momentum of the particles. For high-multiplicity pPb events, a clear particle species
dependence of $v_2$ and $v_3$ is observed. For $\pt < 2\GeV$, the $v_2$ and $v_3$ values of $\PKzS$ particles
are larger than those of $\PgL$/$\PagL$ particles at the same \pt. This splitting
effect between two particle species is found to be stronger
in \pPb\ than in \PbPb\ collisions in the same multiplicity range.
When divided by the number of constituent quarks and compared at the same transverse
kinetic energy per quark, both $v_2$ and $v_3$ for $\PKzS$ particles are observed
to be consistent with those for $\PgL$/$\PagL$ particles at the 10\% level in \pPb\ collisions.
This consistency extends over a wide range of particle transverse kinetic energy and event
multiplicities.
}

\hypersetup{%
pdfauthor={CMS Collaboration},%
pdftitle={Long-range two-particle correlations of strange hadrons with charged particles in pPb and PbPb collisions at LHC energies},%
pdfsubject={CMS},%
pdfkeywords={CMS, ridge, long-range, correlations, flow, high-multiplicity}}

\maketitle %maketitle comes after all the front information has been supplied
\section{Introduction}

Studies of multiparticle correlations provide important insights into the underlying
mechanism of particle production in high-energy collisions of protons and nuclei.
A key feature of such correlations in ultrarelativistic nucleus-nucleus (\AonA) collisions
is the observation of a pronounced structure on the near side (relative azimuthal angle
$\abs{\dphi} \approx 0$) that extends over a large range in relative pseudorapidity
($\abs{\deta}$ up to 4 units or more). This feature, known as the ``ridge", has been found over a wide range of \AonA\
energies and system sizes at both the Relativistic Heavy Ion Collider (RHIC)~\cite{Adams:2005ph,Abelev:2009af,Alver:2008gk,Alver:2009id,Abelev:2009jv}
and the Large Hadron Collider (LHC)~\cite{Chatrchyan:2011eka,Chatrchyan:2012wg,Aamodt:2011by,ATLAS:2012at,CMS:2013bza}
and is interpreted as arising primarily from the collective hydrodynamic flow of a
strongly interacting, expanding medium~\cite{Ollitrault:1992bk,Alver:2010gr}.

Similar
long-range correlations have also been discovered in proton-proton (\Pp\Pp)~\cite{Khachatryan:2010gv},
proton-lead (\pPb)~\cite{CMS:2012qk,alice:2012qe,atlas:2012fa}, and deuteron-gold (\dAu)~\cite{Adare:2014keg}
collisions with high final-state particle multiplicity.
As the collision volume size is reduced, it is possible that the system
will not be able to equilibrate and the hydrodynamic description will
break down.  As such, there has been no consensus on the origin of the particle correlation structure
in these smaller systems.
A variety of theoretical models have
been proposed to interpret this phenomenon in \Pp\Pp\ \cite{Li:2012hc}, \pPb, and \dAu\ collisions. Besides hydrodynamic effects in a high-density
system~\cite{Bozek:2011if,Bozek:2012gr}, an alternate model including gluon saturation in
the incoming nucleons has also been shown to describe these data~\cite{Dusling:2012wy,Dusling:2012cg}.

In hydrodynamical descriptions, the collective flow manifests itself as an azimuthal
anisotropy in the distribution of final-state particles. An additional key consequence
of these models is that the measured anisotropies will depend on the mass of the
particle~\cite{Huovinen:2001cy,Kolb:2003dz,Shen:2011eg}. More specifically, for
particles with transverse momentum below about 2\GeV, the anisotropy will be larger
for lighter particles. The presence of this mass ordering is well established in \AonA\
collisions at RHIC and LHC energies~\cite{STAR,PHENIX,Adler:2003kt,Adler:2001nb,Abelev:2014pua}. This phenomenon
has recently also been observed in \pPb~\cite{ABELEV:2013wsa} and \dAu ~\cite{Adare:2014keg}
collisions, consistent with expectations from hydrodynamic models~\cite{Werner:2013ipa,Bozek:2013ska}.
The analysis presented in this paper aims to further explore this effect by extracting anisotropies
of identified strange mesons (\PKzS) and baryons (\PgL\ and \PagL) in \pPb\ and in \PbPb\
collisions that produce similar final-state particle multiplicity.

The azimuthal correlations of emitted particle pairs are typically characterized
by their Fourier components, $\frac{\rd{}N^\text{pair}}{\rd\Delta\phi} \propto 1 + \sum_{n} 2V_{n\Delta} \cos (n\Delta\phi)$,
where $V_{n\Delta}$ are the two-particle Fourier coefficients and $v_n = \sqrt{V_{n\Delta}}$ denote
the single-particle anisotropy harmonics~\cite{Voloshin:1994mz}. In particular,
the second and third Fourier components are known as elliptic ($v_2$) and triangular ($v_3$) flow, respectively~\cite{Alver:2010gr}.
In hydrodynamical models, $v_2$ and $v_3$ are directly related to the response of the medium to the initial collision geometry and
its fluctuations~\cite{Alver:2010dn,Schenke:2010rr,Qiu:2011hf}. As such, these Fourier components can provide insight into the fundamental transport
properties of the medium.

In \AonA\ collisions at RHIC, a scaling of
$v_2$ as a function of \pt\ with the number of constituent quarks ($n_{q}$) has been
observed in the range $2<\pt<6\GeV$~\cite{Adams:2003am}. Specifically, the values of
$v_2$/$n_{q}$ are found to be very similar for all mesons ($n_{q}=2$) and baryons
($n_{q}=3$) when compared at the same value of $\pt$/$n_{q}$. This empirical scaling
may indicate that final-state hadrons are formed through recombination of quarks in
this \pt\ regime~\cite{Molnar:2003ff,Greco:2003xt,Fries:2003vb}, possibly providing
evidence of deconfinement of quarks and gluons in these systems. At lower \pt\ ($\pt<2\GeV$),
a similar scaling behavior is observed, although, according to perfect fluid hydrodynamics, $v_2$/$n_{q}$ values must be compared
at the same transverse kinetic energy per constituent quark ($\ket$/$n_{q}$,
where $\ket = \sqrt{\smash[b]{m^2 + \pt^2}} - m$) to account for the mass difference of hadrons~\cite{Abelev:2007qg,Adare:2006ti}.

This paper presents an analysis of
two-particle correlations with identified strange hadrons, \PKzS\ and \PgL/\PagL,
in \pPb\ collisions at a center-of-mass energy per nucleon pair (\rootsNN) of
5.02\TeV. With the implementation of a
dedicated high-multiplicity trigger, the 2013 \pPb\ data sample gives access
to multiplicities comparable to those in semi-peripheral \PbPb\ collisions.
Two-particle correlation functions are constructed by associating a \PKzS\ or \PgL/\PagL\ particle
with a charged particle (pairs of \PKzS\ or
\PgL/\PagL\ particles are not studied due to their limited
statistical precision).
In the context of hydrodynamic models, Fourier coefficients of dihadron correlations can be factorized into products of single-particle azimuthal anisotropies. Assuming that this relationship holds, $v_{2}$ and $v_{3}$ are extracted from long-range
two-particle correlations as a function of strange hadron \pt\ and event
multiplicity. To examine the validity of constituent quark number scaling, $v_{2}/n_{q}$ and $v_{3}/n_{q}$ are
obtained as a function of $\ket$/$n_{q}$ for both \PKzS\ and \PgL/\PagL\ particles.
A direct comparison of the \pPb\ and \PbPb\ results over a broad range of similar
multiplicities is presented.

\section{The CMS experiment and data sample}

A description of the CMS detector in the LHC at CERN can be found in Ref.~\cite{JINST}. The
main detector component used in this paper is the tracker,
located in a superconducting solenoid of 6\unit{m} internal diameter,
providing a magnetic field of 3.8\unit{T}. The tracker consists of
1440 silicon pixel and 15\,148 silicon strip detector modules, covering
the pseudorapidity range $\abs{\eta}<2.5$.  For hadrons with $\pt \approx
1\GeV$ and $\abs{\eta} \approx 0$, the impact parameter (distance of closest
approach from the primary collision vertex) resolution is
approximately 100\micron and the \pt\ resolution is 0.8\%.

Also located inside the solenoid are the electromagnetic calorimeter (ECAL)
and the hadron calorimeter (HCAL). The ECAL consists of 75\,848
lead tungstate crystals, arranged in a quasi-projective geometry and distributed in a
barrel region ($\abs{\eta} < 1.48$) and two endcaps that extend to $\abs{\eta} = 3.0$.
The HCAL barrel and endcaps are sampling calorimeters composed of brass and
scintillator plates, covering $\abs{\eta} < 3.0$. Iron/quartz-fiber forward calorimeters (HF) are
placed on each side of the interaction region, covering $2.9<\abs{\eta}<5.2$.
The detailed Monte Carlo (MC) simulation
of the CMS detector response is based on {\sc geant4} \cite{GEANT4}.

The data sample used in this analysis was collected with the CMS detector during the LHC \pPb\ run
in 2013. The total integrated luminosity of the data set is about 35\nbinv~\cite{CMS-PAS-LUM-13-002}.
The beam energies are 4\TeV
for protons and 1.58\TeV per nucleon for lead nuclei, resulting in a center-of-mass
energy per nucleon pair of 5.02\TeV. The direction of the proton beam
was initially set up to be clockwise (20~nb$^{-1}$), and was later reversed (15~nb$^{-1}$). As a result of the energy
difference between the colliding beams, the nucleon-nucleon center-of-mass in the \pPb\
collisions is not at rest with respect to the laboratory frame. Massless particles
emitted at $\eta_\text{cm} = 0$ in the nucleon-nucleon center-of-mass frame will be
detected at $\eta = -0.465$ (clockwise proton beam) or $0.465$ (counterclockwise
proton beam) in the laboratory frame. A sample of peripheral \PbPb\ data at \rootsNN\ = 2.76\TeV corresponding to an integrated luminosity
of about 2.3\mubinv,
collected during the 2011 LHC heavy-ion run, is also analyzed for comparison with \pPb\ data at similar charged-particle
multiplicity ranges.

\section{Online triggering and offline track reconstruction and selection}

The online triggering and the offline reconstruction and selection
follow the same procedure as described in Ref.~\cite{Chatrchyan:2013nka}.
Minimum bias \pPb\ events are triggered by requiring at least one track
with $\pt > 0.4$\GeV to be found in the pixel tracker for a \pPb\ bunch crossing.
Because of hardware limits on the data acquisition rate, only a small fraction
($\sim 10^{-3}$) of all minimum bias triggered events are recorded.
In order to collect a large sample of high-multiplicity \pPb\ collisions, a dedicated
high-multiplicity trigger is also implemented using the CMS Level 1 (L1) and high-level
trigger (HLT) systems. At L1, two event streams were triggered by requiring the total transverse energy summed over ECAL and HCAL to be greater than 20 or 40\GeV. Charged tracks are then reconstructed
online at the HLT using the three layers of pixel detectors, and requiring a track
origin within a cylindrical region of 30\unit{cm} length along the beam and 0.2\unit{cm} radius
perpendicular to the beam. For each event, the number of pixel tracks (${N}_\text{trk}^\text{online}$) with
$\abs{\eta}<2.4$ and $\pt > 0.4\GeV$ is counted separately for each vertex. Only tracks with a distance of closest approach of 0.4\unit{cm} or less to one of the vertices are included. The online selection requires ${N}_\text{trk}^\text{online}$ for the vertex with the most tracks to exceed a specific value. Data are taken with thresholds of ${N}_\text{trk}^\text{online}>100, 130$
(from events with L1 threshold of 20\GeV), and $160, 190$ (from events with L1 threshold of 40\GeV).
While all events with $N^\text{online}_\text{trk}>190$ are
accepted, only a fraction of the events from the other thresholds are
kept.  This fraction is dependent on the instantaneous luminosity. Data from both the minimum bias
trigger and high-multiplicity trigger are retained for offline
analysis.

In the offline analysis, hadronic collisions are selected by the presence of at least one tower with energy above 3\GeV in each of the two HF calorimeters. Events are also required to contain at
least one reconstructed primary vertex within 15\unit{cm} of the nominal interaction point
along the beam axis and within 0.15\unit{cm} transverse to the beam trajectory.
At least two reconstructed tracks are required to be associated with the primary vertex, a condition that is important only for minimum bias events.
Beam related background is suppressed by rejecting events for which less than 25\%
of all reconstructed tracks pass the \textit{high-purity} selection (as defined in Ref.~\cite{Chatrchyan:2014fea}). The \pPb\ instantaneous luminosity
provided by the LHC in the 2013 run resulted in a 3\% probability of
having at least one additional interaction present in the same bunch crossing
(pile-up events). The procedure used for rejecting pile-up events is described in
Ref.~\cite{Chatrchyan:2013nka} and is based on the number of tracks associated with each
reconstructed vertex and the distance between different vertices.
A purity of 99.8\% for single \pPb\ collision events is achieved for the highest
multiplicity \pPb\ interactions studied in this paper. With the selection criteria
above, 97--98\% of the events are found to be selected among those \pPb\ interactions
simulated with the \EPOS\ \textsc{lhc}~\cite{Pierog:2013ria} and \HIJING\ 2.1~\cite{Gyulassy:1994ew}
event generators that have at least one particle from the \pPb\ interaction with energy $E>3$\GeV in each of the $\eta$ ranges $-5<\eta <-3$ and $3<\eta <5$.

In this analysis, \textit{high-purity} tracks are used
to select primary tracks (tracks originating from the \pPb\ interaction). Additional requirements are applied to enhance the
purity of primary tracks. The significance of the separation along the beam
axis ($z$) between the track and the best vertex, $d_z/\sigma(d_z)$, and the significance
of the impact parameter relative to the best vertex transverse to the beam, $d_\mathrm{T}/\sigma(d_\mathrm{T})$,
must be less than 3, and the relative \pt uncertainty,
$\sigma(\pt)/\pt$, must be less than 10\%.
To ensure high tracking efficiency and to reduce the rate of misreconstructed
tracks, primary tracks with $\abs{\eta}<2.4$ and $\pt > 0.3\GeV$ are used in the analysis
(a \pt\ cutoff of 0.4\GeV is used in the multiplicity determination to match the HLT requirement). Based on simulation studies using {\sc geant4} to propagate
particles from the \HIJING\ event generator,
the combined geometrical acceptance and efficiency for primary
track reconstruction exceeds 60\% for $\pt \approx 0.3$\GeV and $\abs{\eta}<2.4$. The efficiency
is greater than 90\% in the $\abs{\eta}<1$ region for $\pt>0.6\GeV$. For the event
multiplicity range studied in this paper, no dependence of the tracking efficiency on multiplicity
is found and the rate of misreconstructed tracks is 1--2\%.

The entire \pPb\ data set is divided into classes based on the reconstructed track multiplicity, \noff,
where primary tracks with $\abs{\eta}<2.4$ and $\pt >0.4$\GeV are counted. Details of the
multiplicity classification in this analysis, including the fraction of the full multiplicity distribution and the
average number of primary tracks before and after correcting for detector effects in
each multiplicity range, are provided in Ref.~\cite{Chatrchyan:2013nka}.

A subset of semi-peripheral \PbPb\ data collected during the 2011 LHC heavy-ion run with a minimum bias
trigger are also reanalyzed in order to directly compare \pPb\ and \PbPb\ systems
at the same collision multiplicity. The reanalyzed events were in the range of 50--100\% centrality, where centrality is defined as the fraction of the total inelastic cross section, with 0\% denoting the most central collisions. This sample was reprocessed using the same
event selection and track reconstruction algorithm as for the present \pPb\ analysis. A description of the 2011 \PbPb\ data can be found in
Refs.~\cite{Chatrchyan:2012xq,Chatrchyan:2013nka}.

\section{Reconstruction of \texorpdfstring{\PKzS\ and \PgL/\PagL}{K0S and Lambda/anti-Lambda} candidates}
\label{sec:V0}

The reconstruction technique for \PKzS\ and \PgL/\PagL\ candidates (generally referred to as $V^{0}$s)
at CMS was first described in Ref.~\cite{Khachatryan:2011tm}. To increase the efficiency for tracks with low momentum and large impact parameters,
both characteristic of the \PKzS\ and \PgL/\PagL\ decay products, the standard \textit{loose}
selection of tracks (as defined in Ref.~\cite{Chatrchyan:2014fea}) is used in reconstructing the \PKzS\ and \PgL/\PagL\ candidates.
Oppositely charged tracks with at least 4 hits and both transverse and longitudinal impact parameter significances
greater than 1 (with respect to the primary vertex) are first selected to form a secondary vertex. The distance of closest approach
of the pair of tracks is required to be less than 0.5\unit{cm}. The fitted vertex in $x$, $y$, $z$ of each pair of
tracks is required to have a $\chi^{2}$ value normalized by the number of degrees of freedom less than 7.
The pair of tracks is assumed to be $\Pgpp\Pgpm$
in \PKzS\ reconstruction, while the assumption of $\Pgpm\Pp(\Pgpp\Pap)$ is used in
\PgL\ (\PagL) reconstruction. For \PgL/\PagL, the lower-momentum track is assumed to
be the pion.

Due to the long lifetime of \PKzS\ and \PgL/\PagL\ particles,
a requirement on the significance of the $V^{0}$ decay length, which is the three-dimensional
distance between the primary and $V^{0}$ vertices divided by its uncertainty, to be greater than 5
is applied to
reduce background contributions. To remove \PKzS\ candidates misidentified as \PgL/\PagL\ particles and vice
versa, the \PgL/\PagL\ (\PKzS) candidates must have a corresponding $\Pgpp\Pgpm(\Pp\Pgpm)$
mass more than 20 (10)\MeV away from the PDG value of the \PKzS\ (\PgL) mass~\cite{Beringer:1900zz}.
The angle $\theta^{\text{point}}$ between the $V^0$ momentum vector and
the vector connecting the primary and $V^0$ vertices
is required to satisfy $\cos\theta^{\text{point}}>0.999$.  This reduces the
effect of nuclear interactions, random combinations of tracks, and
\PgL/\PagL\ particles originating from weak decays of
$\Xi$ and $\PgOm$ particles. From MC simulations using {\GEANTfour} and the \HIJING\ event generator, it is found that
the contribution of \PgL/\PagL\ particles from weak
decays is less than 3\% after this requirement.
The \PKzS\ (\PgL/\PagL) reconstruction efficiency is about 6\%
(1\%) for $\pt \approx 1\GeV$ and 20\% $(10\%)$ for $\pt>3\GeV$ within $\abs{\eta}<2.4$.  This
efficiency includes the effects of acceptance and the branching ratio for $V^0$ particle decays into
neutral particles. The relatively low reconstruction efficiency of the $V^{0}$ candidates
is primarily due to the decay length cut and the low efficiency for reconstructing daughter tracks with $\pt<0.3$\GeV or
large impact parameters.

\begin{figure*}[thb]
\centering
\includegraphics[width=\linewidth]{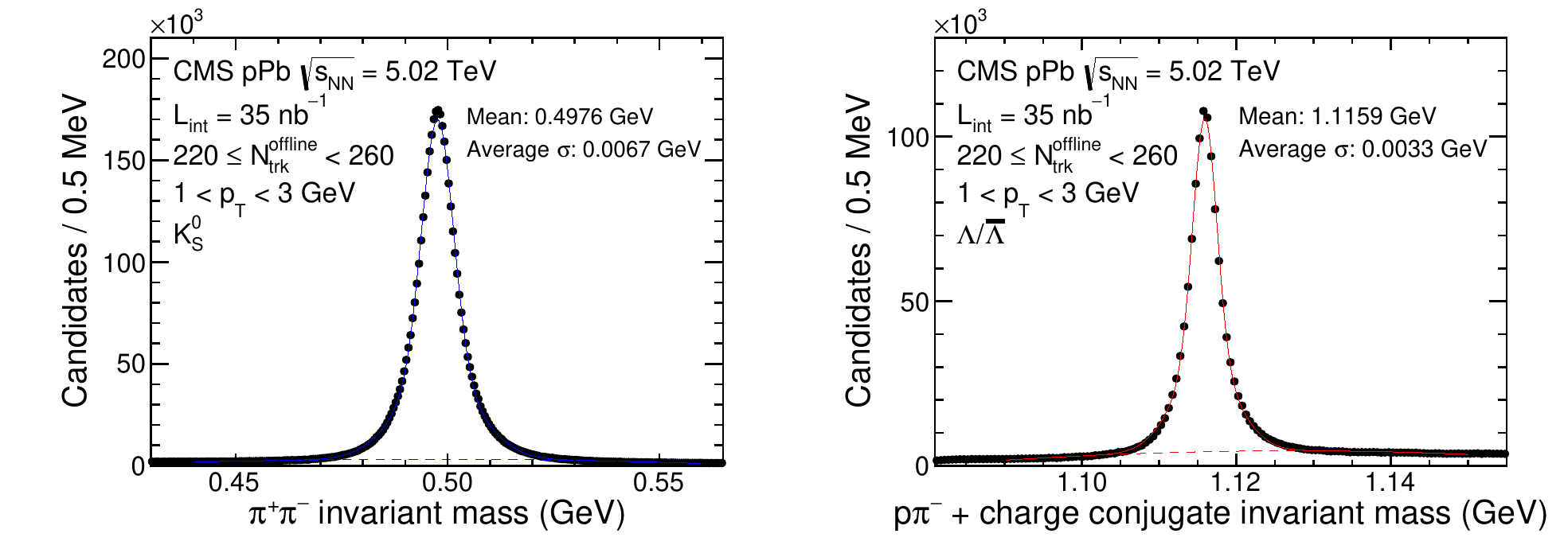}
  \caption{ \label{fig:v0mass} Invariant mass distribution of \PKzS\ (left) and \PgL/\PagL\ (right)
  candidates in the \pt\ range of 1--3\GeV for $220 \leq \noff < 260$ in
  \pPb\ collisions at \rootsNN\ = 5.02\TeV. The solid line shows the fit function of
  a double Gaussian plus a 4th-order polynomial (dashed line).
   }
\end{figure*}

Examples of invariant mass distributions of reconstructed \PKzS\ and \PgL/\PagL\
candidates are shown in Fig.~\ref{fig:v0mass} for \pPb\ data, with $V^0$ \pt\ in the
range of 1--3\GeV and event multiplicity in the range $220 \leq \noff < 260$. Since the results for \PgL\ and \PagL\ are found to be consistent, they have been combined in this analysis. The $V^0$ peaks can be clearly
identified with little background. The true $V^0$ signal peak is well described by a double
Gaussian function (with a common mean), while the background is modeled by a 4th-order
polynomial function fit over the entire mass range shown in Fig.~\ref{fig:v0mass}. The mass window of $\pm2\sigma$ wide around the center
of the peak is defined as the ``peak region'', where $\sigma$ represents
the root mean square of the two standard deviations of the double Gaussian functions weighted by the yields (with typical value of $\sigma$
indicated in Fig.~\ref{fig:v0mass}). To estimate the contribution
of background candidates in the peak region to the correlation measurement,
a ``sideband region'' is chosen that includes $V^0$ candidates from outside the $\pm$3$\sigma$ mass range
around the $V^{0}$ mass to the limit of the mass distributions
shown in Fig.~\ref{fig:v0mass}.

\section{Analysis of two-particle correlations}

The construction of the two-particle correlation function follows the same
procedure established in Refs.~\cite{Chatrchyan:2011eka,Chatrchyan:2012wg,CMS:2012qk,Chatrchyan:2013nka}.
However, in this paper, reconstructed $V^{0}$ candidates from either
the peak or sideband region are taken as ``trigger'' particles within
a given \pttrg\ range, instead of charged tracks as used in previous publications.
The number of trigger $V^{0}$ candidates in the event is denoted by $N_\text{trig}$.
Particle pairs are formed by associating each trigger particle with the remaining
charged primary tracks in a specified \ptass\ interval (which can be either the same as
or different from the \pttrg\ range). The two-dimensional (2D) correlation function is defined in the same way as in previous analyses as
\begin{equation}
\label{2pcorr_incl}
\frac{1}{N_\text{trig}}\frac{\rd^{2}N^\text{pair}}{\rd\Delta\eta\, \rd\Delta\phi}
= B(0,0)\times\frac{S(\Delta\eta,\Delta\phi)}{B(\Delta\eta,\Delta\phi)},
\end{equation}
where $\Delta\eta$ and $\Delta\phi$ are the differences in $\eta$
and $\phi$ of the pair. The same-event pair distribution, $S(\Delta\eta,\Delta\phi)$,
represents the yield of particle pairs normalized by $N_\text{trig}$ from the same event,
\begin{equation}
\label{eq:signal}
S(\Delta\eta,\Delta\phi) = \frac{1}{N_\text{trig}}\frac{\rd^{2}N^\text{same}}{\rd\Delta\eta\, \rd\Delta\phi}.
\end{equation}
The mixed-event pair distribution,
\begin{equation}
\label{eq:background}
B(\Delta\eta,\Delta\phi) = \frac{1}{N_\text{trig}}\frac{\rd^{2}N^\text{mix}}{\rd\Delta\eta\, \rd\Delta\phi},
\end{equation}
is constructed by pairing the trigger $V^{0}$ candidates in each
event with the associated charged primary tracks from 20 different randomly selected
events in the same 2\unit{cm} wide range of vertex position in the $z$ direction and from the same track multiplicity class.
Here, $N^\text{mix}$ denotes the number of pairs taken from the mixed events.
The ratio $B(0,0)/B(\Delta\eta,\Delta\phi)$ mainly accounts for the pair acceptance
effects, with $B(0,0)$ representing the mixed-event associated yield for
both particles of the pair going in approximately the same direction and
thus having maximum pair acceptance (with a bin width of 0.3 in $\Delta\eta$ and
$\pi/16$ in $\Delta\phi$). Thus, the quantity in Eq.~(\ref{2pcorr_incl}) is effectively the per-trigger-particle associated yield. A pair is removed if the associated particle belongs
to a daughter track of any trigger $V^0$ candidate (this contribution is negligible since
associated particles are mostly primary tracks).

The same-event and mixed-event pair distributions are first calculated for each event, and
then averaged over all the events within the track multiplicity class. The range
of $0<\abs{\deta}<4.8$ and $0<\abs{\dphi}<\pi$ is used to fill one quadrant of the
($\Delta\eta,\Delta\phi$) histograms, with the other three quadrants filled (for
illustration purposes) by reflection to cover a ($\Delta\eta,\Delta\phi$) range of
$-4.8<\deta<4.8$ and $-\pi/2<\dphi<3\pi/2$ for the 2D correlation functions, as
will be shown later in Fig.~\ref{fig:Corr_Fcn_Fig1}. In performing the correlation
analyses, each reconstructed primary track and $V^{0}$ candidate is weighted by a
correction factor, following the procedure described in
Refs.~\cite{Chatrchyan:2011eka,Chatrchyan:2012wg,CMS:2012qk,Chatrchyan:2013nka}.
This correction is also applied in calculating $N_\text{trig}$.
This factor accounts for detector effects including the reconstruction efficiency,
the detector acceptance, and the fraction of misreconstructed tracks.
This correction factor is found to have a negligible effect on the azimuthal anisotropy
harmonics.

\subsection{Extraction of \texorpdfstring{$v_n$}{v[n]} harmonics}

Motivated by hydrodynamic models of long-range correlations in \pPb\ collisions,
azimuthal anisotropy harmonics of \PKzS\ and \PgL/\PagL\ particles are extracted via
a Fourier decomposition of \dphi\ correlation functions averaged over $\abs{\deta}>2$ (to
remove short-range correlations such as jet fragmentation),
\begin{linenomath}
\begin{equation}
\label{eq:Vn}
\frac{1}{N_\text{trig}}\frac{\rd N^\text{pair}}{\rd\Delta\phi} = \frac{N_{\text{assoc}}}{2\pi} \left[ 1+\sum\limits_{n} 2V_{n\Delta} \cos (n\Delta\phi)\right],
\end{equation}
\end{linenomath}
as was done in Refs.~\cite{Chatrchyan:2011eka,Chatrchyan:2012wg,CMS:2012qk,Chatrchyan:2013nka}.
Here, $V_{n\Delta}$ are the Fourier coefficients and $N_{\text{assoc}}$
represents the total number of pairs per trigger $V^{0}$ particle for a given
$(\pttrg, \ptass)$ bin. The first three Fourier terms are included in the fits to
the correlation functions. Including additional terms has a negligible effect on the
results of the Fourier fit.

If the observed two-particle azimuthal correlations arise purely as the result
of convoluting anisotropic distributions of single particles, then the $V_{n\Delta}$
coefficients can be factorized into the product of single-particle anisotropies~\cite{Chatrchyan:2013nka},
\begin{linenomath}
\begin{equation}
\label{eq:factorization}
V_{n\Delta}(\pttrg,\ptass)=v_{n}(\pttrg) \times v_{n}(\ptass).
\end{equation}
\end{linenomath}
Following this assumption, the elliptic $(v_2)$ and triangular $(v_3)$
anisotropy harmonics of $V^{0}$
particles can be extracted as a function of \pt\ from the fitted Fourier coefficients,
\begin{linenomath}
\begin{equation}
\label{eq:Vnpt}
v_{n}(\pt^{V^{0}}) = \frac{V_{n\Delta}(\pt^{V^{0}},\ptref)}{\sqrt{V_{n\Delta}(\ptref,\ptref)}},\qquad n=2, 3.
\end{equation}
\end{linenomath}
\noindent Here, a fixed \ptref\ range for the ``reference'' charged primary
particles is chosen to be $0.3<\pt<3.0$\GeV (the lowest \pt region accessible by CMS and the same as was used in Ref.~\cite{Chatrchyan:2013nka}),
to minimize correlations from back-to-back jets at higher \pt.

The $v_n$ values are first extracted for $V^0$ candidates from the peak region
(which contains small contributions from background $V^0$s) and sideband region,
denoted as \vnobs\ and \vnbkg, respectively. The $v_n$ signal of true $V^0$ particles is
denoted by \vnsig\ and is obtained by
\begin{linenomath}
\begin{equation}
\label{eq:v2sig}
\vnsig\ = \frac{\vnobs-(1-f^\text{sig}) \times \vnbkg }{f^\text{sig}},\qquad n=2, 3,
\end{equation}
\end{linenomath}
assuming \vnsig\ and \vnbkg\ are independent from each other. Here,
$f^\text{sig}$ represents the signal yield fraction in the peak region determined
by the fits to the mass distribution shown in Fig.~\ref{fig:v0mass}. This fraction
exceeds 80\% for \PgL/\PagL\ candidates at $\pt>1$\GeV and is above 95\% for \PKzS\ candidates over
the entire \pt\ range.

\subsection{Systematic uncertainties}

Table~\ref{tab:syst-table} summarizes different sources of systematic uncertainties
in \vnsig\ (identical for \PKzS\ and \PgL/\PagL\ particles) for \pPb\ and \PbPb\ data. The dominant sources of systematic uncertainties are related to the reconstruction
of $V^{0}$ candidates. The systematic effects are found to have no
dependence on \pt\ so the estimated systematic uncertainties
are assumed to be constant percentages over the entire \pt\ range. Systematic uncertainties in \vtrdsig\ are assumed to be the same
as those in \vsecsig, as was done in Ref.~\cite{Chatrchyan:2013nka}.

The range of the $V^{0}$ mass distributions used in fitting the signal plus background (Fig.~\ref{fig:v0mass}) is varied by 10\%. This change, which could affect the value of $f^\text{sig}$ used in Eq.~(\ref{eq:v2sig}), yields a systematic uncertainty of less than 1\% for the \vsecsig\ results. Changing the mass range included in the peak region could impact the values of both $f^\text{sig}$ and \vsecobs. For a variation from ${\pm}1\sigma$ to ${\pm}3\sigma$, the \vsecsig\ values are found to be consistent within 2\%. Systematic uncertainties due to selection of different sideband mass regions, which could change \vsecbkg, are estimated to be 2.2\%.
Possible contamination by residual misidentified
$V^{0}$ candidates (\ie, \PKzS\ as \PgL/\PagL, and vice versa) is also investigated.
Variation of the invariant mass range used
to reject misidentified $V^0$ candidates leads to variations of less
than 2\% on \vsecsig. Systematic effects related to
selection of the $V^{0}$ candidates are evaluated by varying the requirements
on the decay length significance and $\cos\theta^{\text{point}}$, resulting in
an uncertainty of 3\%.
As misalignment of the tracker detector elements can affect the $V^0$
reconstruction performance, an alternative detector geometry is studied. Compared to the standard configuration, this alternative has the two halves of the barrel pixel detector shifted in opposite directions along the beam by a distance on the order of 100\mum. The values of \vsecsig\ found using the shifted configuration differed by less than 2\% from the default ones.

To test the procedure of extracting the $V^{0}$ signal $v_2$
from Eq.~(\ref{eq:v2sig}), a study using \EPOS\ {\textsc{lhc}\xspace} \pPb\ MC events is performed to compare
the extracted \vsecsig\ results with the generator-level \PKzS\ and \PgL/\PagL\ values.
The agreement is found to be better than 4\%.
Other systematic uncertainties introduced by the high-multiplicity trigger
efficiency (1\%) and possible residual pile-up effects (1--2\%) for \pPb\ data are estimated
in the same way as in Ref.~\cite{Chatrchyan:2013nka}, and found to make only a small contribution.
The various sources of systematic uncertainties are added together in quadrature to arrive
at the final systematic uncertainties (6.9\% for \pPb\ and 6.6\% for \PbPb), which are shown as
shaded boxes in Figs.~\ref{fig:v2_PID_lowN}--\ref{fig:v3_PID_highN}.

\begin{table*}[ht]
\topcaption{\label{tab:syst-table} Summary of systematic uncertainties in \vnsig\ for \pPb\ and \PbPb\ data.}

\begin{center}
\begin{tabular}{lcc}
 Source			&	 \pPb (\%)		&	\PbPb (\%) \\
\hline
 $V^0$ mass distribution range used in fit 	&   1	&	1	\\
 Size of $V^0$ mass region for signal	  &  2   &	2	\\
 Size and location of $V^0$ mass sideband region 	&  2.2	&	2.2	\\
 Misidentified $V^0$ mass region	& 2	&	2	\\
 $V^0$ selection criteria		&	3	&	3	\\
 Tracker misalignment		&	2	&	2	\\
 MC closure	test			&	4	&	4	\\
 Trigger efficiency			&	2	&	---	\\
 Pile-up			&	1	&	---	\\
\hline
 Total		&	6.9	&	6.6	\\
\end{tabular}
\end{center}
\end{table*}

\section{Results}

\begin{figure*}[thb]
\centering
\includegraphics[width=0.8\linewidth]{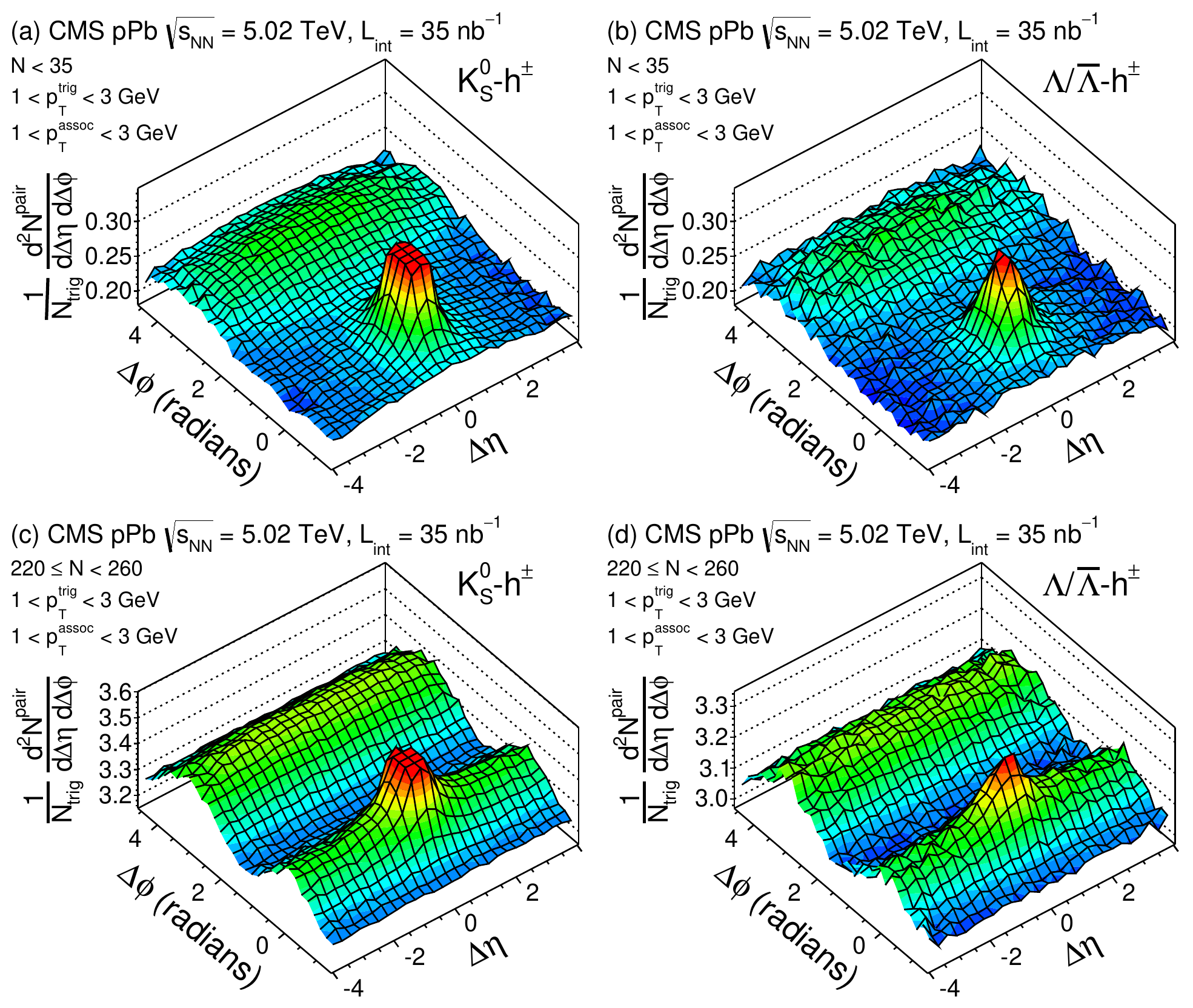}
  \caption{ \label{fig:Corr_Fcn_Fig1} The 2D two-particle correlation functions
  in \pPb\ collisions at \rootsNN\ = 5.02\TeV for pairs of a \PKzS\ (a,c) or \PgL/\PagL\ (b,d)
  trigger particle and a charged associated particle ($h^{\pm}$), with $1<\pttrg<3\GeV$
  and $1<\ptass<3\GeV$, in the multiplicity ranges $\noff< 35$ (a, b) and
  $220 \leq \noff< 260$ (c, d). The sharp near-side peak from jet correlations
  is truncated to emphasize the structure outside that region.
   }
\end{figure*}

The 2D two-particle correlation functions measured in \pPb\ collisions for
pairs of a \PKzS\ (left) and \PgL/\PagL\ (right) trigger particles
and a charged associated particle ($h^{\pm}$) are shown in Fig.~\ref{fig:Corr_Fcn_Fig1}
in the \pt\ range of 1--3\GeV. The 2D correlation functions are corrected
for the background $V^{0}$ candidates, following the same approach of correcting $v_n$ in Eq.~(\ref{eq:v2sig}).
The correction is negligible in this \pt\ range because
of the high signal yield fraction of $V^{0}$ candidates.
For low-multiplicity events ($\noff < 35$, Figs.~\ref{fig:Corr_Fcn_Fig1} (a) and (b)),
a sharp peak near $(\deta, \dphi) = (0, 0)$ due to jet fragmentation (truncated
for better illustration of the full correlation structure) can be clearly observed
for both $\PKzS$--$h^{\pm}$ and $\PgL/\PagL$--$h^{\pm}$ correlations. Moving to
high-multiplicity events ($220 \leq \noff< 260$, Figs.~\ref{fig:Corr_Fcn_Fig1}
(c) and (d)), in addition to the peak from jet fragmentation,
a pronounced long-range structure is seen at $\dphi \approx 0$, extending at least 4.8 units
in $\abs{\deta}$. This structure was previously observed in high-multiplicity ($\noff \sim 110$)
\Pp\Pp\ collisions at \roots\ = 7\TeV~\cite{Khachatryan:2010gv} and \pPb collisions
at \rootsNN\ = 5.02\TeV~\cite{CMS:2012qk,alice:2012qe,atlas:2012fa,Chatrchyan:2013nka}
for inclusive charged particles, and also for identified charged pions, kaons, and
protons in \pPb collisions at \rootsNN\ = 5.02\TeV~\cite{ABELEV:2013wsa}. A similar long-range
correlation structure has also been extensively studied in \AonA collisions over a
wide range of energies~\cite{Adams:2005ph,Abelev:2009af,Alver:2008gk,Alver:2009id,Abelev:2009jv,Chatrchyan:2011eka,Chatrchyan:2012wg,Aamodt:2011by,ATLAS:2012at},
where it is believed to arise primarily from collective flow of a strongly interacting
medium~\cite{Voloshin:1994mz}.

\begin{figure*}[thb]
\centering
\includegraphics[width=0.8\linewidth]{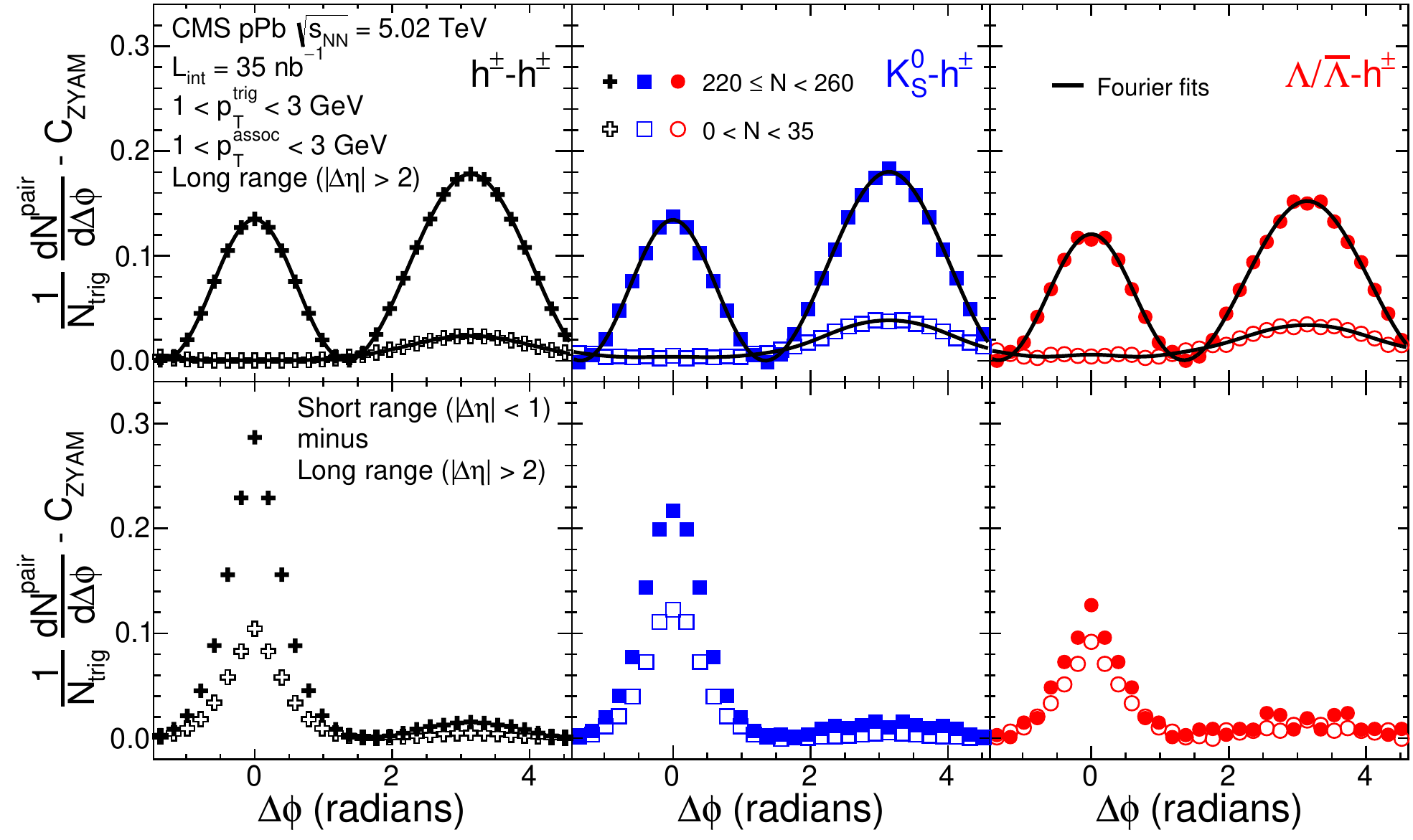}
  \caption{ \label{fig:Corr_Proj_Fig2}
     The 1D \dphi\ correlation functions from \pPb\ data after applying the ZYAM procedure,
     in the multiplicity range $\noff\ < 35$ (open) and $220 \leq \noff\ < 260$ (filled),
     for trigger particles composed of inclusive charged particles (left), \PKzS\ particles
     (middle), and \PgL/\PagL\ particles (right).
     Selection of a fixed \pttrg\ and \ptass\ range of both 1--3\GeV      is shown for the long-range region ($\abs{\deta}>2$) on top and the short-range ($\abs{\deta}<1$) minus long-range region on the bottom.
     The curves on the top panels correspond to the Fourier fits including the first three terms. Statistical uncertainties
     are smaller than the size of the markers.
   }
\end{figure*}

To investigate the correlation structure for different species of particles in
detail, one-dimensional (1D) distributions in $\Delta\phi$ are found
by averaging the signal and mixed-event 2D distributions over $\abs{\deta} < 1$ (defined
as the ``short-range region'') and $\abs{\deta} > 2$ (defined as the ``long-range region''),
as done in Refs.~\cite{Khachatryan:2010gv,Chatrchyan:2011eka,Chatrchyan:2012wg,CMS:2012qk,Chatrchyan:2013nka}.
Fig.~\ref{fig:Corr_Proj_Fig2} shows the
1D $\Delta\phi$ correlation functions from \pPb\ data for trigger
particles composed of inclusive charged particles (left)~\cite{Chatrchyan:2013nka}, \PKzS\
particles (middle), and \PgL/\PagL\ particles (right), in
the multiplicity range $\noff\ < 35$ (open) and $220 \leq \noff\ < 260$
(filled). The curves show the Fourier fits from Eq.~(\ref{eq:Vn}) to the long-range region,
which will be discussed in detail later. Following the standard zero-yield-at-minimum (ZYAM) procedure~\cite{Chatrchyan:2013nka}, each distribution is shifted to have zero
associated yield at its minimum to represent the correlated portion of the associated yield. Selection of fixed \pttrg\ and \ptass\ ranges of 1--3\GeV is shown for the long-range region (top) and for the difference of the short- and long-range regions (bottom) in Fig.~\ref{fig:Corr_Proj_Fig2}.
As illustrated in Fig.~\ref{fig:Corr_Fcn_Fig1}, the near-side long-range
signal remains nearly constant in
$\Delta\eta$. Therefore, by taking a difference of 1D \dphi\ projections between
the short- and long-range regions, the near-side jet correlations can be extracted.
As shown in the bottom panels of Fig.~\ref{fig:Corr_Proj_Fig2},
due to biases in multiplicity selection toward higher \pt\ jets, a larger jet peak yield is observed
for events selected with higher multiplicities. Because charged particles are directly used in determining the multiplicity in the event,
this selection bias is much stronger for charged particles than \PKzS\ and \PgL/\PagL\ hadrons.
For $\noff\ < 35$, no near-side correlations are observed in the long-range region for
any particle species. The \PbPb\ data show qualitatively the same behavior as the \pPb\ data, and thus are not presented here.

\begin{figure*}[thb]
\centering
\includegraphics[width=0.8\linewidth]{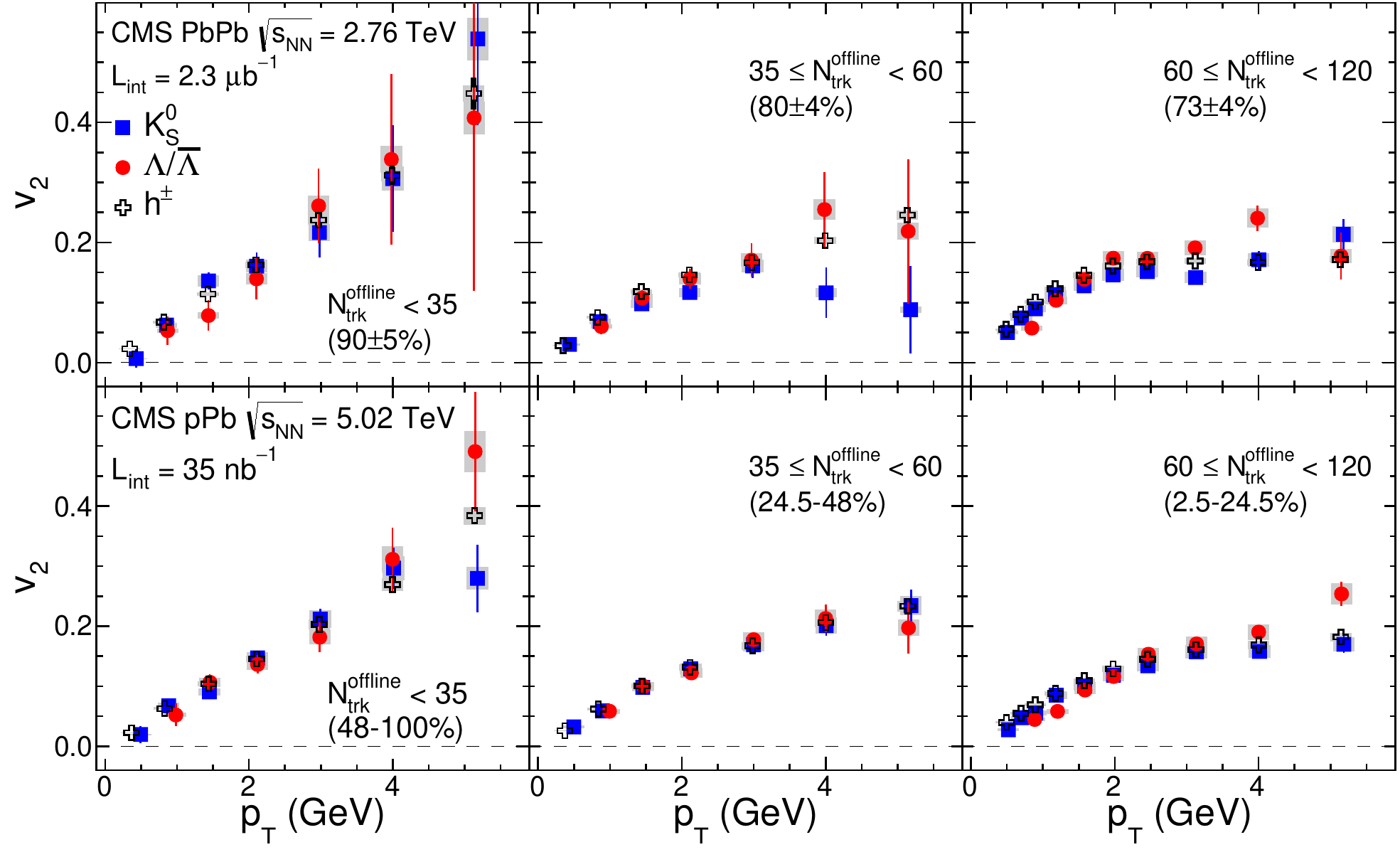}
  \caption{ \label{fig:v2_PID_lowN}
    The $v_2$ results for \PKzS\ (filled squares) and \PgL/\PagL\ (filled circles) particles
    as a function of \pt\
    for three multiplicity ranges obtained from minimum bias triggered \PbPb\
    sample at \rootsNN\ = 2.76\TeV (top row) and \pPb\ sample at \rootsNN\ = 5.02\TeV
    (bottom row). The error bars correspond to statistical uncertainties,
    while the shaded areas denote the systematic uncertainties.
    The values in parentheses give the mean and standard deviation of the
    HF fractional cross section for \PbPb\ and the range of the fraction of the full multiplicity distribution included for \pPb.
   }
\end{figure*}

Recently, the $v_2$ anisotropy harmonics for charged pions, kaons, and protons
have been studied using two-particle correlations in \pPb\ collisions~\cite{ABELEV:2013wsa},
and are found to be qualitatively consistent with hydrodynamic models~\cite{Werner:2013ipa,Bozek:2013ska}.
In this paper, the elliptic ($v_2$) and triangular ($v_3$) flow harmonics
of \PKzS\ and \PgL/\PagL\ particles are extracted from the Fourier decomposition of 1D
$\Delta\phi$ correlation functions for the long-range region ($|\Delta\eta| > 2$)
in a significantly larger sample of \pPb\ collisions such that the particle
species dependence of $v_n$ can be investigated in detail. In Fig.~\ref{fig:v2_PID_lowN},
the $v_2^{\text{sig}}$ of \PKzS\ and \PgL/\PagL\ particles
are plotted as a function of \pt\ for the three lowest multiplicity ranges in \PbPb\
and \pPb\ collisions. These data were recorded
using a minimum bias trigger. The range of the fraction of the
full multiplicity distribution that each multiplicity selection corresponds to, as determined in
Ref.~\cite{Chatrchyan:2013nka}, is also specified in the figure. In contrast to most
other \PbPb\ analyses, the present work uses multiplicity to classify events, instead of
the total energy deposited in HF (the standard procedure of
centrality determination in \PbPb)~\cite{Chatrchyan:2012xq,Chatrchyan:2013nka}.
By examining the HF energy distribution for \PbPb\ events in each of the multiplicity ranges, the corresponding
average HF fractional cross section (and its standard deviation) can be determined, which are
presented for \PbPb\ data in the figure.

In the low multiplicity region (Fig.~\ref{fig:v2_PID_lowN}),
the $v_2$ values of \PKzS\ and \PgL/\PagL\ particles are compatible within
statistical uncertainties. As there is no evident long-range near-side correlation
seen in these low-multiplicity events, the extracted $v_2$ most likely reflects
back-to-back jet correlations on the away side. Away-side jet correlations
typically appear as a peak structure around $\dphi \approx \pi$,
which contributes to various orders of Fourier terms.

\begin{figure*}[thb]
\centering
\includegraphics[width=\linewidth]{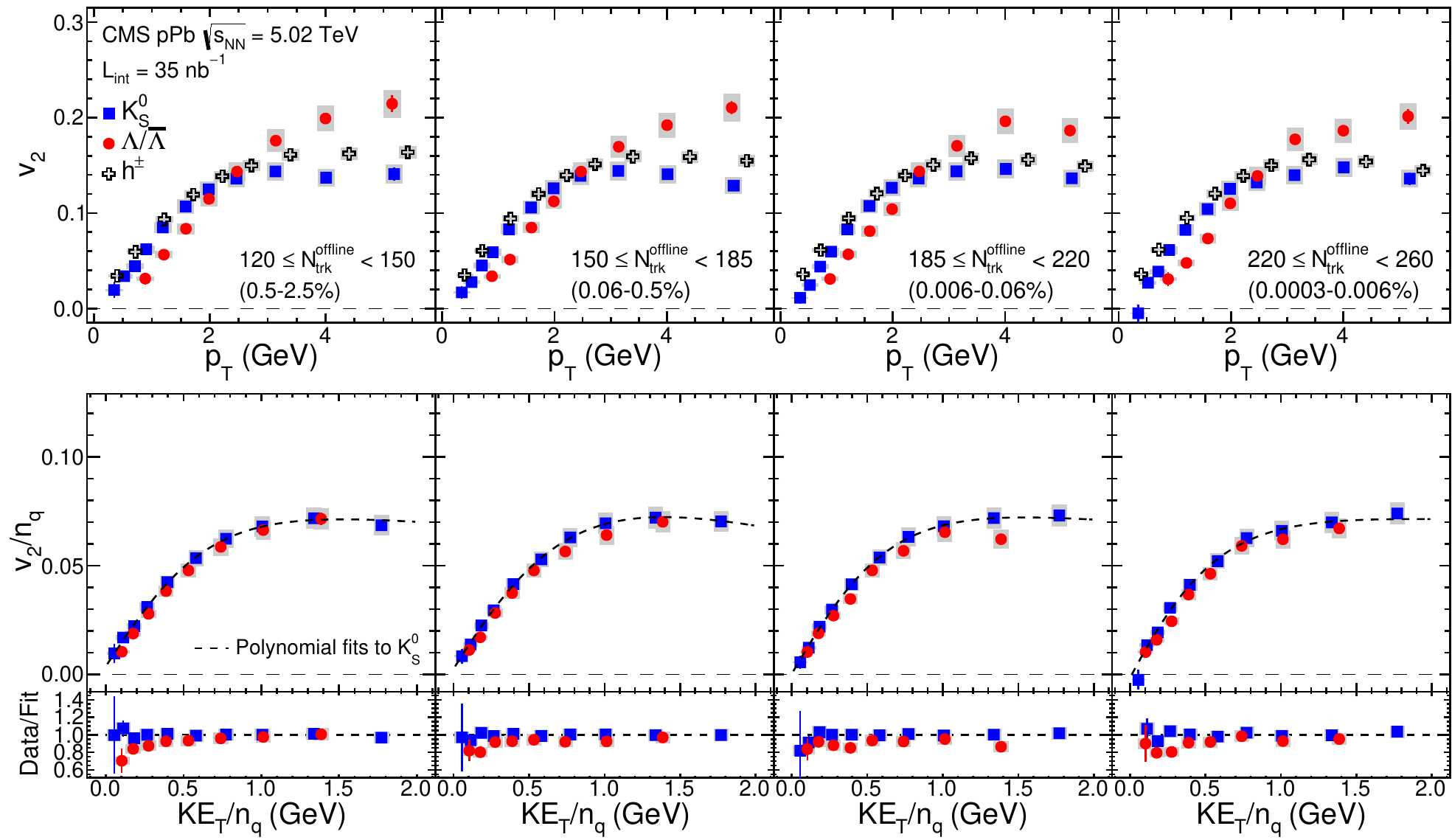}
  \caption{ \label{fig:v2_PID_highN_pPb}
    Top row: the $v_2$ results for \PKzS\ (filled squares), \PgL/\PagL\ (filled circles),
    and inclusive charged particles (open crosses) as a function of \pt\
    for four multiplicity ranges obtained from high-multiplicity triggered \pPb\ sample at \rootsNN\ = 5.02\TeV.
    Middle row: the $v_2/n_q$ ratios for \PKzS\ (filled squares) and \PgL/\PagL\ (filled circles) particles
    as a function of $\ket/n_q$, along with a fit to the \PKzS\ results using a polynomial function.
    Bottom row: ratios of $v_2/n_q$ for \PKzS\ and \PgL/\PagL\ particles to
    the fitted polynomial function as a function of $\ket/n_q$.
    The error bars correspond to statistical uncertainties, while the shaded areas denote
    the systematic uncertainties. The values in parentheses give the range of the
    fraction of the full multiplicity distribution included for \pPb.
   }
\end{figure*}

The top row of Fig.~\ref{fig:v2_PID_highN_pPb} shows the measured $v_2$ values for \PKzS\ and \PgL/\PagL\ particles as a function of \pt\ from the high multiplicity \pPb\ data, along with the previously published results for inclusive charged particles~\cite{Chatrchyan:2013nka}.
In the $\pt\lesssim 2\GeV$ region
for all high-multiplicity ranges, the $v_2$ values of \PKzS\ particles are larger than those for \PgL/\PagL\ particles at
each \pt value. Both of them are consistently below the $v_2$ values of inclusive
charged particles. As most charged particles are pions,
the data indicate that lighter particle species exhibit a stronger azimuthal anisotropy signal.
This mass ordering behavior is consistent with expectations in hydrodynamic models and the observation
in 0--20\% centrality \pPb\ collisions~\cite{ABELEV:2013wsa}. A similar trend was first observed
in \AonA\ collisions at RHIC~\cite{Adler:2003kt,Adler:2001nb}. At higher \pt, the $v_2$ values of \PgL/\PagL\ particles are larger than those of \PKzS.
The inclusive charged particle $v_2$ values fall between the values of the
two identified strange hadron species but are much closer to the $v_2$ values for \PKzS\ particles.
Note that the ratio of baryon to meson yield in \pPb\ collisions is enhanced at higher \pt,
an effect that becomes stronger as multiplicity increases~\cite{Chatrchyan:2013eya,Abelev:2013haa}.
This should also be taken into account when comparing $v_n$ values between inclusive and identified particles. Comparing the results in Fig.~\ref{fig:v2_PID_lowN} and Fig.~\ref{fig:v2_PID_highN_pPb}, the dependence of $v_2$ on the particle species may already be emerging in the multiplicity range of $60 \leq \noff < 120$.

The scaling behavior of $v_2$ divided by the number of constituent quarks as a function of
transverse kinetic energy per quark, $\ket/n_{q}$, is investigated for
high-multiplicity \pPb\ events in the middle row of Fig.~\ref{fig:v2_PID_highN_pPb}.
After scaling by the number of quarks, the $v_2$ distributions for
\PKzS\ and \PgL/\PagL\ particles are found to be in agreement. The
middle row of Fig.~\ref{fig:v2_PID_highN_pPb} also shows the result of fitting a polynomial
function to the \PKzS\ data. The bottom row of Fig.~\ref{fig:v2_PID_highN_pPb} shows the
$n_q$-scaled $v_2$ results for \PKzS\ and \PgL/\PagL\ particles divided by
this polynomial function fit, indicating that the scaling is valid to
better than 10\% over most of the $\ket/n_{q}$ range,
except for $\ket/n_{q}<0.2$\GeV where the deviation grows to about 20\%.
In \AonA\ collisions, this approximate scaling behavior is conjectured to be related to quark
recombination~\cite{Molnar:2003ff,Greco:2003xt,Fries:2003vb}, which postulates
that collective flow is developed among constituent quarks before they combine
into final-state hadrons. Note that the scaling of $v_2$ with the number of constituent quarks was originally observed as a function
of \pt, instead of \ket, for the intermediate \pt\ range of a few \GeV~\cite{Adams:2003am},
and interpreted in a simple picture of quark coalescence~\cite{Molnar:2003ff}.
However, it was later discovered that when plotted as a function of \ket\ in order to
remove the mass difference of identified hadrons, the scaling appears to hold over the entire
kinematic range~\cite{Abelev:2007qg,Adare:2006ti}. However, this scaling
behavior is not expected to be exact at low \pt\ in hydrodynamic models because of the impact of radial flow.
As the $v_n$ data tend to approach a constant value as a function of \pt\ or \ket\ for \pt\ $\gtrsim$ 2\GeV, the scaling behavior
in terms of \pt\ and \ket\ cannot be differentiated in that regime. Therefore, the $n_{q}$-scaled $v_n$ results in this paper
are presented as a function of $\ket/n_{q}$ in order to explore the scaling behavior over a wider kinematic range.

\begin{figure*}[thb]
\centering
\includegraphics[width=\linewidth]{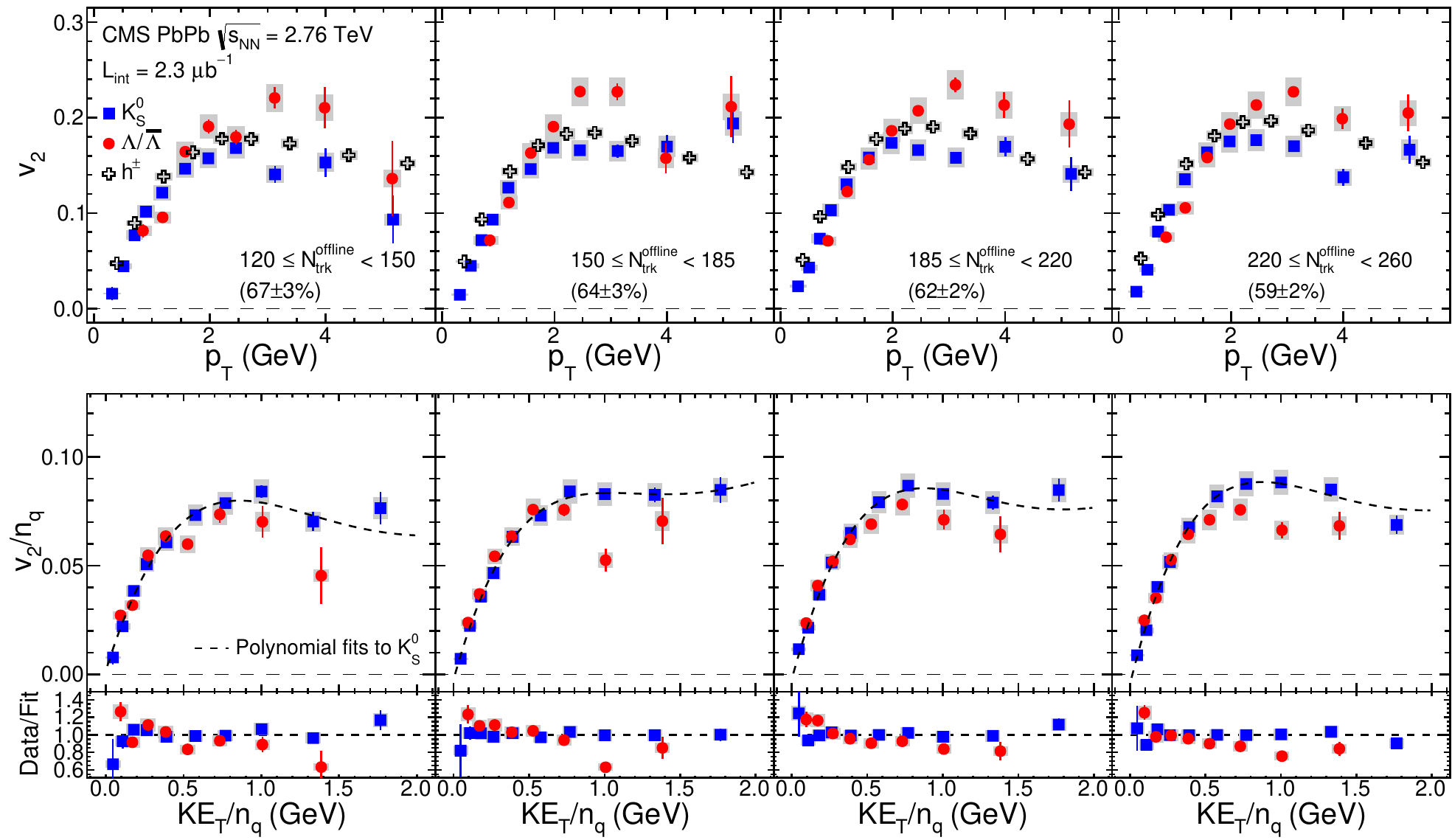}
  \caption{ \label{fig:v2_PID_highN_PbPb}
    Top row: the $v_2$ results for \PKzS\ (filled squares), \PgL/\PagL\ (filled circles),
    and inclusive charged particles (open crosses) as a function of \pt\
    for four multiplicity ranges obtained from minimum bias triggered \PbPb\ sample at \rootsNN\ = 2.76\TeV.
    Middle row: the $v_2/n_q$ ratios for \PKzS\ (filled squares) and \PgL/\PagL\ (filled circles) particles
    as a function of $\ket/n_q$. Bottom row: ratios of $v_2/n_q$ for \PKzS\ and \PgL/\PagL\ particles to
    a smooth fit function of $v_2/n_q$ for \PKzS\ particles as a function of $\ket/n_q$.
    The error bars correspond to statistical uncertainties, while the shaded areas denote
    the systematic uncertainties. The values in parentheses give the mean and standard deviation of the HF fractional cross section
    for \PbPb.
   }
\end{figure*}

The particle species dependence of $v_2$ and its scaling behavior is also studied
in \PbPb\ data over the same multiplicity ranges as for
the \pPb\ data, as shown in Fig.~\ref{fig:v2_PID_highN_PbPb}. The mean and standard
deviation of the HF fractional cross section of the \PbPb\ data are indicated on the plots.
Qualitatively, a similar particle-species dependence of $v_2$
is observed. However, the mass ordering effect is found to be less
evident in \PbPb\ data than in \pPb\ data for all multiplicity ranges. In hydrodynamic models, this
may indicate a stronger radial flow is developed in the \pPb\ system as its energy density is higher
than that of a \PbPb\ system due to having a smaller size system at the same multiplicity.
Moreover, the $n_{q}$-scaled $v_2$ data in \PbPb\ at similar multiplicities suggest
a stronger violation of constituent quark number scaling, up to 25\%, than is observed in \pPb,
especially for higher $\ket/n_{q}$ values. This is also observed in peripheral \AuAu\ collisions
at RHIC, while the scaling applies more closely for central \AuAu\ collisions~\cite{Adare:2012vq}.

\begin{figure*}[thb]\centering
\centering
\includegraphics[width=0.45\linewidth]{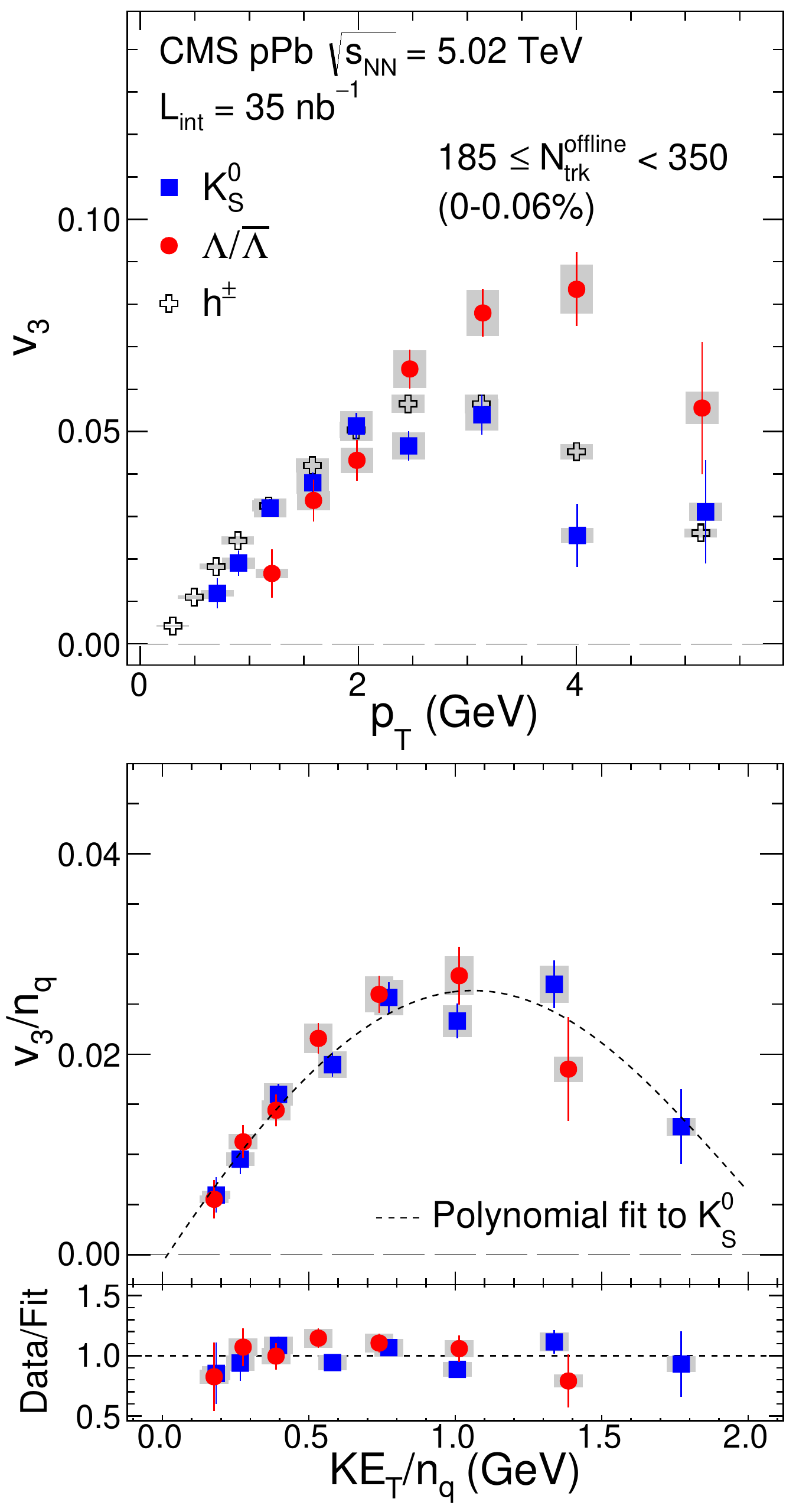}
\includegraphics[width=0.45\linewidth]{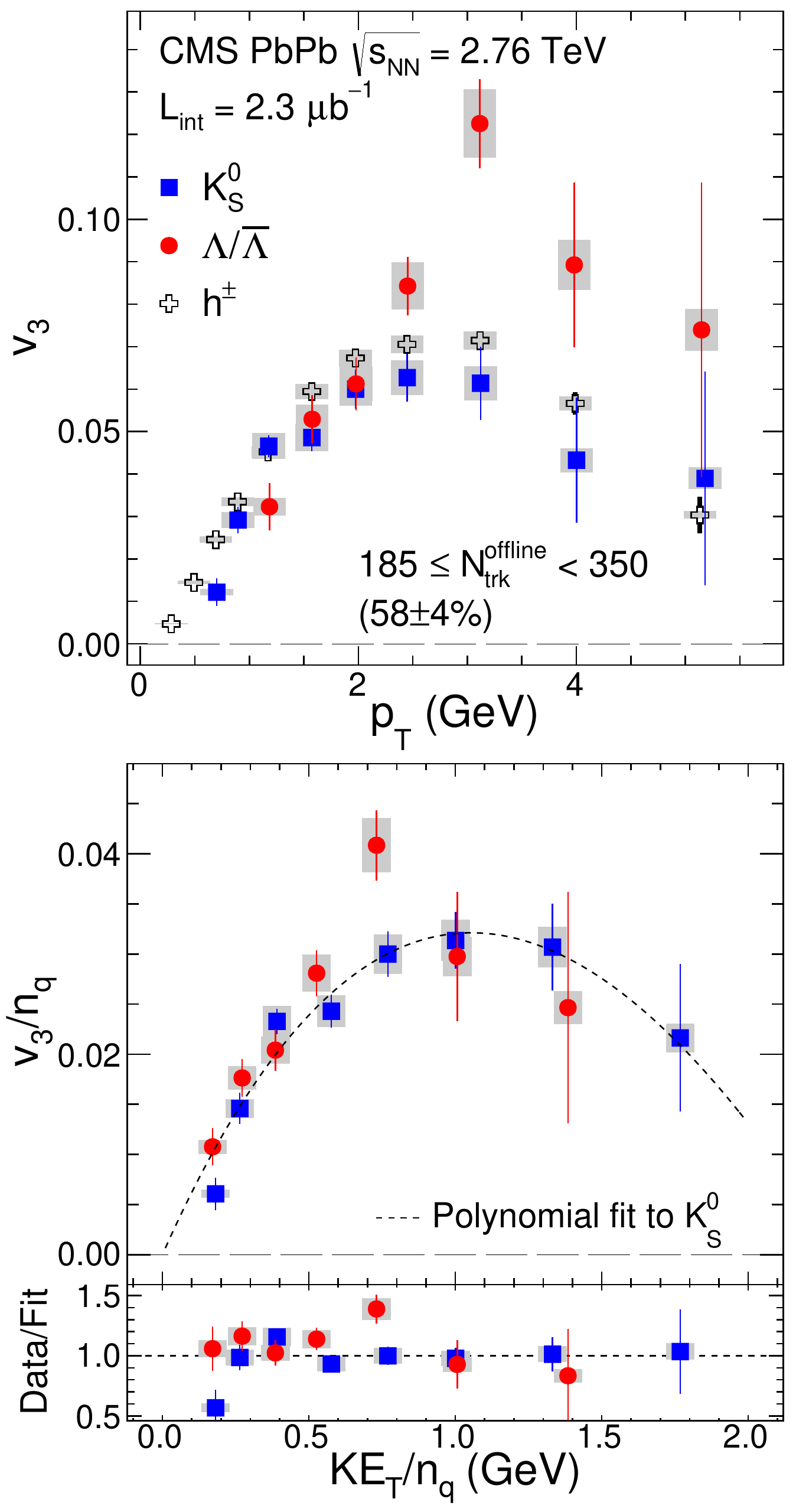}
  \caption{ \label{fig:v3_PID_highN}
    Top: the $v_3$ results for \PKzS\ (filled squares), \PgL/\PagL\ (filled circles),
    and inclusive charged particles (open crosses) as a function of \pt\ for
    the multiplicity range $185 \leq \noff < 350$ in \pPb\ collisions at
    \rootsNN\ = 5.02\TeV (left) and in \PbPb\ collisions at \rootsNN\ = 2.76\TeV (right).
    Bottom: the $n_{q}$-scaled $v_3$ values of \PKzS\ (filled squares)
    and \PgL/\PagL\ (filled circles) particles as a function of $\ket/n_q$ for the same two systems. Ratios of $v_n/n_q$ to
    a smooth fit function of $v_n/n_q$ for \PKzS\ particles as a function of $\ket/n_q$ are also shown.
    The error bars correspond to statistical uncertainties, while the shaded areas denote
    the systematic uncertainties. The values in parentheses give the mean and standard deviation of the HF fractional cross section
    for \PbPb\ and the range of the fraction of the full multiplicity distribution included for \pPb.
   }
\end{figure*}

The triangular flow harmonic, $v_3$, of \PKzS\ and \PgL/\PagL\ particles is also extracted in \pPb\
and \PbPb\ collisions, as shown in Fig.~\ref{fig:v3_PID_highN}. Due to limited statistical
precision, only the result in the multiplicity range $185 \leq \noff < 350$ is presented.
A similar species dependence of $v_3$ to that of $v_2$ is observed and,
within the statistical uncertainties, the $v_3$ values scaled by the constituent quark number
for \PKzS\ and \PgL/\PagL\ particles match at the level of 20\% over the full $\ket/n_{q}$ range.
To date, no calculations of the quark number scaling of triangular flow, $v_3$, have been performed in the
parton recombination model.

\section{Summary}

Measurements of two-particle
correlations with an identified \PKzS\ or \PgL/\PagL\ trigger particle have been presented over a broad
transverse momentum and
pseudorapidity range in \pPb\ collisions at \rootsNN\ = 5.02\TeV and \PbPb\
collisions at \rootsNN\ = 2.76\TeV. With the implementation of a high-multiplicity
trigger during the LHC 2013 \pPb\ run, the identified particle correlation data
in \pPb\ collisions are explored over a broad particle multiplicity range,
comparable to that covered by 50--100\% centrality \PbPb\ collisions.
The long-range ($\abs{\deta}>2$) correlations are quantified in terms of azimuthal
anisotropy Fourier harmonics ($v_n$) motivated by hydrodynamic models.
In low-multiplicity \pPb\ and \PbPb\ events, similar $v_2$ values of \PKzS\ and \PgL/\PagL\ particles
are observed, which likely originate from back-to-back jet correlations.
For higher event multiplicities, a particle species dependence of $v_2(\pt)$ and $v_3(\pt)$
is observed. For $\pt \lesssim 2$\GeV,
the values of $v_n$ for \PKzS\ particles are found to be larger than those of \PgL/\PagL\ particles, while this order
is reversed at higher \pt. This behavior is consistent with RHIC and LHC results in
\AonA\ collisions and for identified charged hadrons in \pPb\ and \dAu\ collisions.
For similar event multiplicities, the particle species dependence of $v_2$ and $v_3$ at low \pt\
is observed to be more pronounced in \pPb\ than in \PbPb\ collisions. In the context of hydrodynamic models,
this may indicate that a stronger radial flow boost is developed in \pPb\ collisions.
Furthermore, constituent quark number scaling of $v_2$ and $v_3$
between \PKzS\ and \PgL/\PagL\ particles is found to apply for \PbPb\ and high-multiplicity
\pPb\ events. The constituent quark number scaling is found to hold at the 10\% (25\%) level in \pPb\ (\PbPb) collisions, for similar event multiplicities. It will be interesting to see if this scaling law continues to hold for other particles.
The results presented in this paper
provide important input to the further exploration of the possible collective flow origin
of long-range correlations, and can be used to evaluate models of quark recombination in a deconfined medium
of quarks and gluons.

\begin{acknowledgments}
We congratulate our colleagues in the CERN accelerator departments for the excellent performance of the LHC and thank the technical and administrative staffs at CERN and at other CMS institutes for their contributions to the success of the CMS effort. In addition, we gratefully acknowledge the computing centres and personnel of the Worldwide LHC Computing Grid for delivering so effectively the computing infrastructure essential to our analyses. Finally, we acknowledge the enduring support for the construction and operation of the LHC and the CMS detector provided by the following funding agencies: BMWFW and FWF (Austria); FNRS and FWO (Belgium); CNPq, CAPES, FAPERJ, and FAPESP (Brazil); MES (Bulgaria); CERN; CAS, MoST, and NSFC (China); COLCIENCIAS (Colombia); MSES and CSF (Croatia); RPF (Cyprus); MoER, ERC IUT and ERDF (Estonia); Academy of Finland, MEC, and HIP (Finland); CEA and CNRS/IN2P3 (France); BMBF, DFG, and HGF (Germany); GSRT (Greece); OTKA and NIH (Hungary); DAE and DST (India); IPM (Iran); SFI (Ireland); INFN (Italy); NRF and WCU (Republic of Korea); LAS (Lithuania); MOE and UM (Malaysia); CINVESTAV, CONACYT, SEP, and UASLP-FAI (Mexico); MBIE (New Zealand); PAEC (Pakistan); MSHE and NSC (Poland); FCT (Portugal); JINR (Dubna); MON, RosAtom, RAS and RFBR (Russia); MESTD (Serbia); SEIDI and CPAN (Spain); Swiss Funding Agencies (Switzerland); MST (Taipei); ThEPCenter, IPST, STAR and NSTDA (Thailand); TUBITAK and TAEK (Turkey); NASU and SFFR (Ukraine); STFC (United Kingdom); DOE and NSF (USA).

Individuals have received support from the Marie-Curie programme and the European Research Council and EPLANET (European Union); the Leventis Foundation; the A. P. Sloan Foundation; the Alexander von Humboldt Foundation; the Belgian Federal Science Policy Office; the Fonds pour la Formation \`a la Recherche dans l'Industrie et dans l'Agriculture (FRIA-Belgium); the Agentschap voor Innovatie door Wetenschap en Technologie (IWT-Belgium); the Ministry of Education, Youth and Sports (MEYS) of the Czech Republic; the Council of Science and Industrial Research, India; the HOMING PLUS programme of Foundation for Polish Science, cofinanced from European Union, Regional Development Fund; the Compagnia di San Paolo (Torino); the Consorzio per la Fisica (Trieste); MIUR project 20108T4XTM (Italy); the Thalis and Aristeia programmes cofinanced by EU-ESF and the Greek NSRF; and the National Priorities Research Program by Qatar National Research Fund.
\end{acknowledgments}

\bibliography{auto_generated}   % will be created by the tdr script.
\clearpage \cleardoublepage \appendix\section{The CMS Collaboration \label{app:collab}}\begin{sloppypar}\hyphenpenalty=5000\widowpenalty=500\clubpenalty=5000\textbf{Yerevan Physics Institute,  Yerevan,  Armenia}\\*[0pt]
V.~Khachatryan, A.M.~Sirunyan, A.~Tumasyan
\vskip\cmsinstskip
\textbf{Institut f\"{u}r Hochenergiephysik der OeAW,  Wien,  Austria}\\*[0pt]
W.~Adam, T.~Bergauer, M.~Dragicevic, J.~Er\"{o}, C.~Fabjan\cmsAuthorMark{1}, M.~Friedl, R.~Fr\"{u}hwirth\cmsAuthorMark{1}, V.M.~Ghete, C.~Hartl, N.~H\"{o}rmann, J.~Hrubec, M.~Jeitler\cmsAuthorMark{1}, W.~Kiesenhofer, V.~Kn\"{u}nz, M.~Krammer\cmsAuthorMark{1}, I.~Kr\"{a}tschmer, D.~Liko, I.~Mikulec, D.~Rabady\cmsAuthorMark{2}, B.~Rahbaran, H.~Rohringer, R.~Sch\"{o}fbeck, J.~Strauss, A.~Taurok, W.~Treberer-Treberspurg, W.~Waltenberger, C.-E.~Wulz\cmsAuthorMark{1}
% ...
\vskip\cmsinstskip
% ...
1:~~Also at Vienna University of Technology, Vienna, Austria\\
2:~~Also at CERN, European Organization for Nuclear Research, Geneva, Switzerland\\
% ...

\end{sloppypar}
\end{document}